\documentclass[aps,prl,twocolumn,epsf,showpacs]{revtex4}
\usepackage{graphicx,subfigure,latexsym,amsfonts,amssymb,natbib}
\usepackage{ulem}
\usepackage{color}

\begin{document}

\title{Aggregation of BiTe Monolayer on Bi$_2$Te$_3$(111) Induced by Diffusion of Intercalated Atoms in van der Waals Gap}
\author{Zhi-Wen Wang$^{1}$}
\author{Wen-Kai Huang$^{1}$}
\author{Kai-Wen Zhang$^{1}$}
\author{Da-Jun Shu$^{1,2}$}
\email{djshu@nju.edu.cn}
\author{Mu Wang$^{1,2}$}
\author{Shao-Chun Li$^{1,2,3}$}
\email{scli@nju.edu.cn}
\affiliation{$^{1}$National Laboratory of Solid State Microstructures and School of Physics, Nanjing University, Nanjing 210093, China}
\affiliation{$^{2}$Collaborative Innovation Center of Advanced Microstructures, Nanjing University, Nanjing 210093, China}
\affiliation{$^{3}$Jiangsu Provincial Key Laboratory for Nanotechnology, Nanjing University, Nanjing 210093, China}

\date{\today}

\begin{abstract}
 We report a post-growth aging mechanism of Bi$_2$Te$_3$(111) films with scanning tunneling microscopy in combination with density functional theory calculation. It is found that a monolayered  structure with a squared lattice symmetry gradually aggregates from surface steps. Theoretical calculations indicate that the van der Waals (vdW) gap not only acts as a natural reservoir for self-intercalated Bi and Te atoms, but also provides them  easy diffusion pathways. Once  hopping  out of the gap, these defective atoms prefer to  develop into a two dimensional BiTe superstructure on the Bi$_2$Te$_3$(111) surface driven by positive energy gain. Considering the common nature of weakly bonding between vdW layers, we expect such  unusual diffusion and aggregation  of the intercalated atoms may be of general importance for most kinds of vdW layered materials.
   \end{abstract}
\pacs{}
\maketitle


 Studies on bismuth chalcogenide (Bi$_2$X$_3$) have long been focused regarding the potential in thermoelectricity \cite{WRIGHT-Nature1958}, and also been revived recently by the discovery of topological insulators (TIs) \cite{Moore.Nature,SCZhang.RevModPhys,Kane.RevModPhys,ZhangHJ-natphys2009,Hasan_Nature2009,ChenYL_Science2009,Hasan-PRL2009,WangG.Adv.mater,NatureCom2016}.
It has been confirmed that most Bi$_2$X$_3$ compounds, such as Bi$_2$Te$_3$, Bi$_2$Se$_3$ and Sb$_2$Se$_3$, are prototypical 3D TIs \cite{ZhangHJ-natphys2009,ChenYL_Science2009,Hasan_Nature2009,Hasan-PRL2009}. Hosting the topological surface states and gapped bulk band structure, Bi$_2$X$_3$ becomes promising in the future applications such as quantum computation and spintronic etc \cite{Moore.Nature,SCZhang.RevModPhys,Kane.RevModPhys}.

In spite of the robustness of the topological surface states against  time-reversal invariant perturbation, temporal stability of the surface or interface still greatly affects the Bi$_2$X$_3$'s performance and application. Intensive studies have been conducted regarding the surface atomistic and band structure evolution upon exposing to various gases including air \cite{Kern-PRL2011,ZhouXJ_PNAS2012, Yan-PRL2013, Edmonds_JPCC2014,KaiwenZhang.PhysRevB,CrystalGrowthDesign2011,JSolidStateChemistry2016}, however it still remains elusive whether and how the Bi$_2$X$_3$ intrinsically evolves in ultrahigh vacuum (UHV). ARPES studies indicated that a long stay in UHV induces an energy band splitting \cite {Kern-PRL2011}. A low-energy ion scattering (LEIS) study confirmed a way of surface degradation of Bi$_2$Se$_3$, i.e., the cleaved surface in UHV undergoes a fast Se evaporation and is thus terminated by Bi bilayer \cite{WuRQ_PRL2013-LEIS}. It was also observed that annealing induces the surface segregation of Bi bilayer \cite {Coelho_NanoLett2013}. Therefore, a direct characterization of surface evolution upon long term exposure in UHV is important, which, however, hasn't been reported yet.

In this Letter, we demonstrate a post-growth aggregation on Bi$_2$Te$_3$(111) surface induced by the self-intercalation and diffusion of Bi and Te atoms in the van der Waals (vdW) gap with scanning tunneling microscopy (STM) in combination with density functional theory (DFT) calculation. We observe that upon long term aging, a monolayered Mosaic pattern, presumably composed of Bi and Te, is gradually formed starting at the step edges of Bi$_2$Te$_3$(111) surface. DFT calculations confirm that Bi atoms prefer to be intercalated in the vdW gap between quintuple layers (QLs) than being adsorbed on the surface, and the difference between the formation energy of Bi and Te intercalated atoms in the gap is smaller than that on the surface. The diffusion barriers of the intercalated atoms in the vdW gap are also comparable to those on the surface. We therefore suggest that the vdW gap acts as a reservoir for the bulk defect Bi and Te atoms with thermally activated diffusion pathways. The lateral segregation of the intercalated atoms from the steps leads to a BiTe monolayered structure on Bi$_2$Te$_3$ (111) surface. Furthermore, elastic relaxation due to the stress mismatch between BiTe monolayer and Bi$_2$Te$_3$ surface forms a Mosaic pattern in the BiTe monolayer. In contrast to the previous studies about the surface stability, which are mainly surface decomposition or reaction \cite{WuRQ_PRL2013-LEIS,Coelho_NanoLett2013}, this work presents a long term effect induced by the bulk intercalated atoms. Since intercalation of atoms commonly exist in vdW gaps \cite{MoS-nl14,InChem16}, it is expected that such a novel mechanism may exist in other layered materials as well, considering the intrinsic nature of vdW interactions.


Thin Bi$_2$Te$_3$ films were grown by molecular beam epitaxy (MBE) in UHV with a Unisoku LT-STM-MBE joint system. The base pressure is $1 \times10^{-10}$ mbar. Bi$_2$Te$_3$(111) films were grown on the Si(111)-$(7 \times7)$ substrate at $\sim$ 260$^\circ$C, and $\sim$2 monolayers (MLs) of Te buffer layer was deposited prior to the Bi$_2$Te$_3$(111) film growth. The as-grown Bi$_2$Te$_3$ film was kept in UHV without any further processing prior to STM scan. STM measurements were performed  at $\sim$80 K. Constant current mode was adopted, and a mechanically polished PtIr tip was applied.

\begin{figure}
  \centering
  \includegraphics[width=7.5cm]{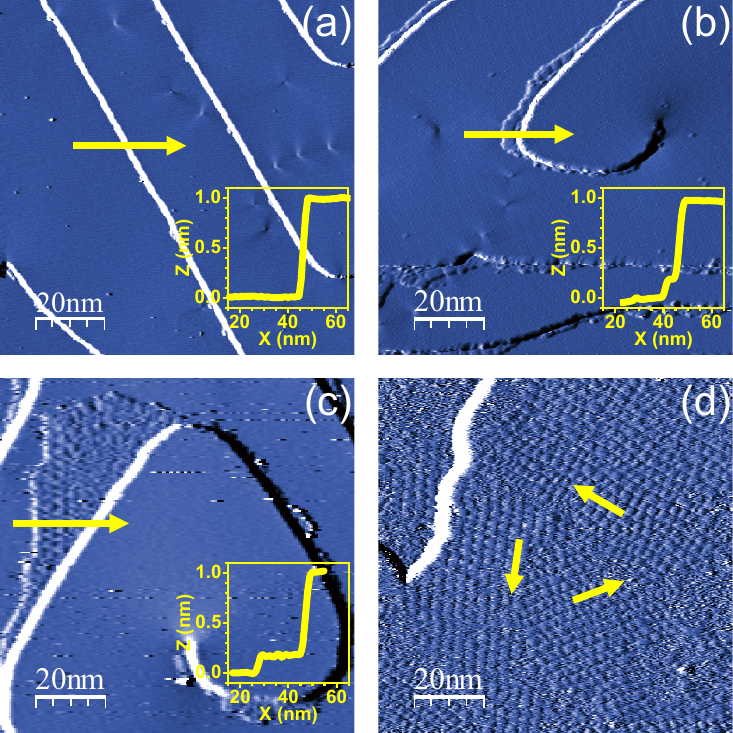}\\
  \caption{(a) Derivative STM image (100 $\times$ 100 nm$^{2}$) of the as grown Bi$_2$Te$_3$(111) surface. $U = +1.0$ $V$, $I$$_t$ $=$ $0.1$ $nA$. (b)-(d) Derivative STM images (100 $\times$ 100 nm$^{2}$) taken after the sample is exposed in UHV for about 4, 6, and 13 days, respectively. $U$ $=$ $+1.0$ $V$, $I$$_t$ $=$ $0.1$ $nA$. Inset plots in (a)-(c) are the line-scan profiles measured along the arrows marked in the corresponding images. Arrows in (d) mark the three equivalent orientations of the superstructure. }\label{fig1}
\end{figure}

Figure\ \ref{fig1} shows the temporal evolution of the surface morphology of Bi$_2$Te$_3$ films upon aging in UHV. The as-grown Bi$_2$Te$_3$ film, as shown in Fig.\ 1(a), is terminated by the spiral steps and atomically flat surface terraces. The step height of Bi$_2$Te$_3$ is $\sim$1 nm, consistent with Bi$_2$Te$_3$ crystal structure and previous report \cite{ZhangT_PRL2009}. Upon long term exposure in UHV, a superstructure starts to develop. Figures\ \ref{fig1}(b) to (d) show the surface morphology after aging in UHV for four days, six days and thirteen days, respectively. The newly formed structure prefers to nucleate adjacent to the Bi$_2$Te$_3$ steps and grows laterally in the lower terrace. The coverage of superstructure increases with the aging time. The thickness of the superstructure is {kept at} $\sim$0.2 nm, as measured in the insets of Fig.\ \ref{fig1}(b) and (c), implying that the superstructure is a monolayered structure. Moreover, as marked by the yellow arrows in Fig.\ \ref{fig1}(d), there  exists  three equivalent domains in the superstructure, oriented $\sim$120$^\circ$  with each other.

\begin{figure}
  \centering
  \includegraphics[width=7.5cm]{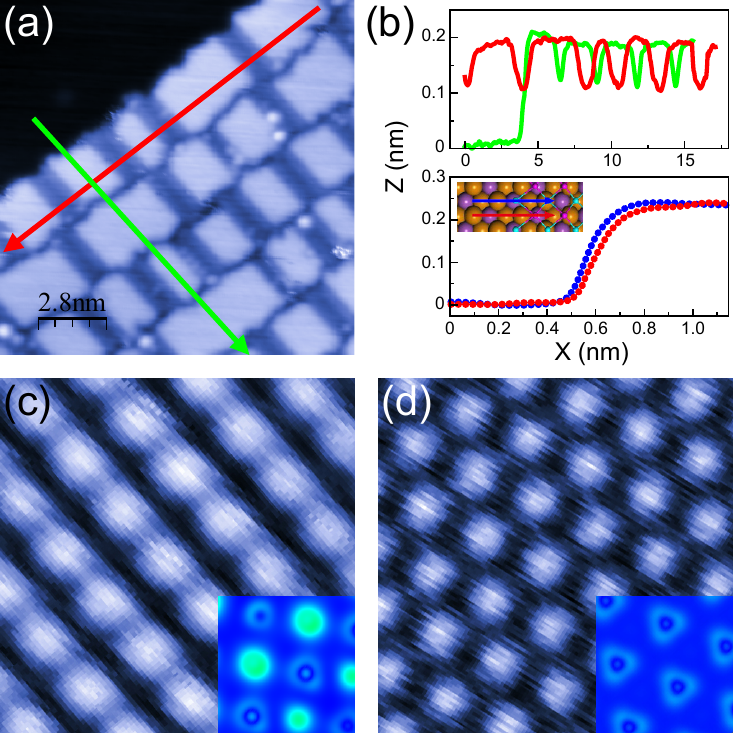}\\
  \caption{(a) STM image (14 $\times$ 14 nm$^{2}$) of the superstructure. The area in the upper-left corner is the  Bi$_2$Te$_3$(111) surface. $U$ $=$ $+1.0$ $V$, $I$$_t$ $=$ $0.1$ $nA$. (b) Line-scan profiles (upper) measured along the arrowed lines marked in (a) and the simulated line-scan profiles (lower) taken along the arrowed lines marked in the inset. The inset shows partial view of the structure of the (4 $\times$ 1) BiTe on a (12 $\times$ 1) supercell of Bi$_2$Te$_3$(111) surface. (c) Atomically resolved STM image (2 $\times$ 2 nm$^{2}$) of the BiTe superstructure. $U$ $=$ $+0.5$ $V$, $I$$_t$ $=$ $0.15$ $nA$. (d) Atomically resolved STM image of the Bi$_2$Te$_3$(111). Insets to (c) and (d) show the corresponding simulated STM images with the tip at 0.1 $\AA$ above the surface.}\label{fig2}
  \end{figure}

Zoom-in images indicate that the superstructure features the ribbon-like structure. The width of the ribbons is distributed uniformly at $\sim$2.5 nm, as illustrated by the green line in Fig.\ \ref{fig2}(a) and the  scan profile in Fig.\ \ref{fig2}(b).  There also  exists  depressions along the ribbons, with a corrugation of $\sim$0.1 nm. The lateral interspace distribution of these depressions varies from $\sim$1.5 nm to $\sim$5.0 nm, as illustrated by the red line in Fig.\ \ref{fig2}(a) and blue scan profile in Fig.\ \ref{fig2}(b). Atomically resolved STM image of the superstructure, Fig.\ \ref{fig2}(c), shows a nearly squared periodicity, different from the hexagonal symmetry in the Bi$_2$Te$_3$(111) surface, Fig.\ \ref{fig2}(d).

It is well known that Bi forms bilayer islands on top of Bi$_2$Te$_3$ surface, with a thickness of $\sim$0.4 nm \cite{,ChenM-APL2012,Coelho_NanoLett2013}, much larger than the superstructure's thickness of $\sim$0.2 nm. Pure Te on Bi$_2$Te$_3$ surface forms islands or clusters near the surface steps \cite{Hoefer-AIPAdv2015}
, in contrast to the observed monolayered superstructure as well. In addition, the symmetry of Bi and Te adlayer structures are both hexagonal  \cite{ChenM-APL2012,Coelho_NanoLett2013,Hoefer-AIPAdv2015}, not in the squared periodicity as observed in Fig.\ \ref{fig2}(c). Therefore, we rule out the possibility that the superstructure is composed of either pure Bi or pure Te. According to the alloy phase diagram of Bi$_m$Te$_n$ \cite{ASM1992}, the plausible candidate with a squared lattice symmetry should be the (001) termination of metastable BiTe phase, which takes a rock-salt structure in a space group $Fm\bar{3}m$ \cite{ASM1992}. Therefore, we propose that the superstructure is composed of both Bi and Te atoms in the BiTe phase.

Since the superstructure forms in the post-growth aging process, it is reasonable to assume that both the required Te and Bi atoms originate within the intercalated atoms from the as-grown Bi$_2$Te$_3$ layers. To verify such a mechanism, we need to check the following questions: (1) {\it How can the atoms be stored as defects in a bulk?} (2) {\it How can the atoms migrate freely to the surface in aging?} To address these issues, we conduct first-principles DFT calculations to compare the formation and diffusion process of Bi and Te atoms in the bulk and on the surface, respectively.

\begin{figure}[t]
 \includegraphics[width=8.0 cm]{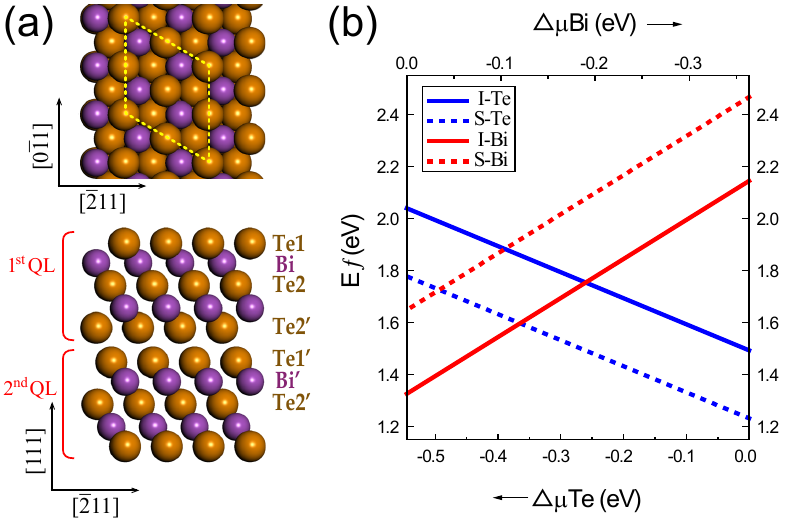}
\caption{ (a) Top view (upper) and side view (lower) of the Bi$_2$Te$_3$(111) surface slab. The yellow dashed lines outline the  $(2\times2)$ supercell adopted in the simulations. (b) The variation of formation energy of Bi (red lines) and  Te  (blue lines) defect atoms with different  chemical potential of  the components, intercalated in the vdW gap (solid lines) or on the Bi$_2$Te$_3$(111) surface (dashed lines).}
\label{fig3}
\end{figure}

The DFT calculations \cite{RevModPhys.64.1045}  were performed with the Vienna Ab Initial Simulation Package (VASP) using the projector augmented wave method \cite{PhysRevB.54.11169,PhysRevB.59.1758}.  The  Perdew-Burke-Ernzerhoff exchange-correlation functional with the semiempirical vdW correction based on Grimmes scheme (PBE-D2) \cite{JCC:JCC20495,JCC:JCC21112,jcp/132/15,PhysRevB.86.184111} was employed,
 with an energy cutoff of 300 eV and no spin-orbit coupling effect included \cite{PhysRevB.86.184111}.
With these settings, the lattice parameters of bulk Bi$_2$Te$_3$  is a = 4.34 $\AA$ and c = 31.71 $\AA$, which agrees well with the experimental values \cite{NAKAJIMA1963479}.  
The Bi$_2$Te$_3$(111) surface was modeled by using a $(2 \times2)$  supercell \cite{PhysRevB.84.144117,Yao2013} consisting of a three-QL slab and a vacuum with thickness of 12 $\AA$, as shown in Fig.\ \ref{fig3}(a).
The surface brillouin zone was sampled with a $\Gamma$-centered $3 \times3$ mesh, which has been tested to be well converged. The positions of atoms in the bottom QL were fixed to mimic the bulk, and the other atoms were relaxed until the forces were converged to 0.01 eV/$\AA$.
A larger $(3 \times 3)$ supercell with a $2 \times2$ k-mesh sampling is used to check the size convergence. The difference is  60 meV  at maximum in  formation energies of Bi  atoms at high-symmetrical sites of the surface and within the vdW gap.
STM images were simulated using the Tersoff and Hamann approximation \cite{PhysRevB.31.1985}.

 As the inter-QL interaction is vdW type, much weaker than the expected covalent bonding within the QLs, it is intuitive to consider the vdW gap as a possible {\it reservoir} to adopt the Te and Bi defect atoms. We thus compare the formation energy of intercalated Te and Bi atoms in the vdW gap and that of adatoms on the surface.
%
 The formation energy of a defect complex containing $m$ Bi and $n$ Te atoms depends on the chemical potential of Bi and Te components. Assuming the approximate equilibrium of  the chemical potential of Bi$_2$Te$_3$  and those of Bi and Te, we obtain the following,
 \begin{equation}
E_{f}(\Delta\mu_{Te})=E_{f}^{0}+\frac{1}{2}m \Delta G_{f}^{0} +(\frac{3}{2} m- n)\Delta\mu_{Te},   \label{eqn:eq4}
\end{equation}
where $E_{f}^{0}=E_{t}^{d}-E_{t}^{h}- m\mu^0_{\mathrm{Bi}} -n \mu^0_{\mathrm{Te}}$, with $E_{t}^{d}$ and $E_{t}^{h}$   the total energy of the Bi$_2$Te$_3$ host  system  with and without the defect atoms, respectively. $\mu^0_{\mathrm{Bi}(\mathrm{Te})}$  are the  free energy per atom in the bulk Bi (Te) metal.    $\Delta G_f^0 =\mu_{Bi_{2}Te_{3}}^{bulk}-2\mu^0_{\mathrm{Bi}}
-3\mu^0_{\mathrm{Te}}$ is the formation energy of  bulk  Bi$_{2}$Te$_{3}$ in reference to the bulk metals.
The  upper and lower bounds of $\Delta \mu_{\mathrm{Te}}$, the chemical potential of Te in reference to the bulk Te metal, are determined by the formation of Te metal and Bi metal, respectively. According to our calculation, $-0.55 \mathrm{eV}< \Delta \mu_{\mathrm{Te}}<0$.


\begin{figure}[t]
 \includegraphics[width=8.0 cm]{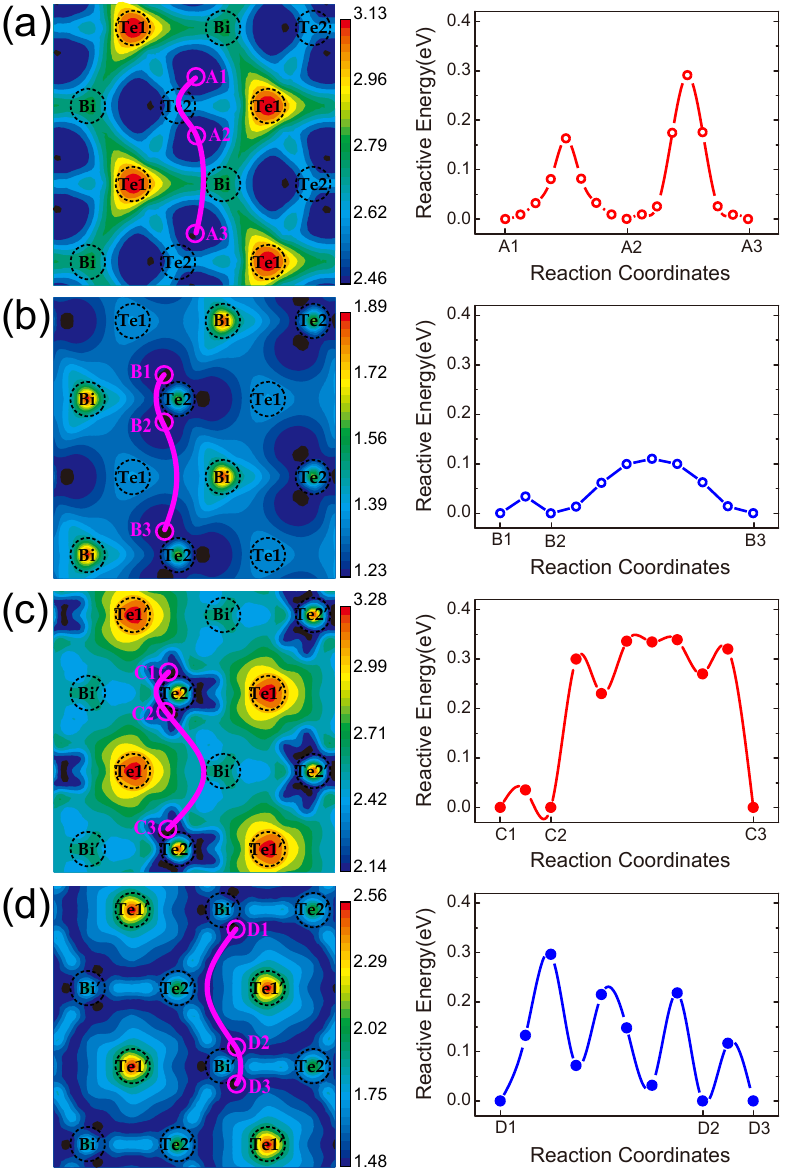}
\caption{Left Panel: Contour plots of the formation energy of (a,c) Bi  and  (b,d) Te defect atoms under the  Te-rich condition. The scale bars for contours are in unit of eV. The defect atoms are (a,b)  on the Bi$_2$Te$_3$(111) surface or (c,d) intercalated in the vdW gap. Black dashed circles schematically show the positions of the host atoms, and pink solid circles correspond to the minimum sites. The pink curved lines schematically show the diffusion pathways of the defect atoms. Right Panel: Energy profiles corresponding to the diffusion pathways of pink curves in the left panel.
}
\label{fig4}
\end{figure}

The potential energy surface (PES) of Bi and Te adatom on the surface in Te-rich limit are plotted  in Fig.\ \ref{fig4}(a) and (b) respectively, $i.e.$, $\Delta \mu_{Te} = 0$.
Similar PES of Bi and Te in the vdW gap are plotted in Fig.\ \ref{fig4}(c) and (d).
Obviously, the formation of Bi adatom is less favorable than that of Te adatom by $\sim$ 1.23 eV. The formation of intercalated Bi  in the vdW gap is also less favorable  than that of Te, but only by $\sim$ 0.65 eV. Note that the minimum sites both on the surface and in the gap deviate evidently from the sites of high symmetry.

It is interesting to note that in Te-rich condition, the formation of Bi defect atom is much easier  in the vdW gap than on the surface, by $\sim$ 0.32 eV. In comparison, formation of Te defect atom is less favorable in the gap than on the surface by $\sim$ 0.26 eV. Therefore the difference between the probability  of forming Bi and Te defect atoms in the gap is quite  smaller than on the surface. This  difference decreases with decreasing $\Delta \mu_{\mathrm{Te}}$, as shown in Fig.\ \ref{fig3}(b).  The intercalation of Bi and Te atoms becomes equally favorable  in the vdW gap at $\Delta\mu_{\mathrm{Te}}$ of $-0.26$ eV.  It means that  the number ratio of Bi and Te intercalated in the gap approaches to 1:1 when the number of Bi adatoms is still much smaller than that of Te adatoms on the surface.

In addition to the formation energy, the diffusion barriers of the defect atoms  are also important because they determine the dynamics during the aging process.  The nudged elastic band (NEB) method \cite{neb95} is used to obtain the accurate  diffusion barrier  for the diffusion pathways based on the PES, as illustrated by the red curves in Fig.\ \ref{fig4}. The limiting barriers for the diffusion of Bi and Te atom on the surface  are obtained to be 0.28  eV (A2-A3) and 0.11 eV (B2-B3), respectively.  Comparatively, the diffusion barriers of Bi and Te  atom in the gap are 0.34 eV (C2-C3) and 0.30 eV (D1-D2), respectively. It is obvious that the diffusion  barriers of both Bi and Te atoms in the vdW gap are still small enough for thermal activation during the aging process.

\begin{figure}[t]
\includegraphics[width=8.0 cm]{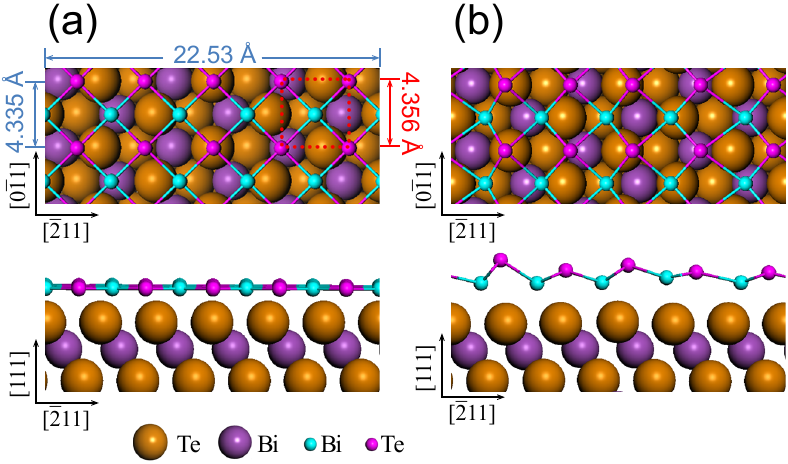}
\caption{ Top views (upper) and side views (lower) of the BiTe monolayer on the  $(6\times1)$ Bi$_2$Te$_3$(111) surface, before (left) and after relaxation (right).
  }
\label{fig5}
\end{figure}

Combining the DFT results and experiments, it follows that a monolayered  Bi$_m$Te$_n$ with $m$:$n$$>$2:3 can be formed as a result of the lateral segregation of intercalated Bi and Te atoms from the vdW gap.  The aforementioned BiTe satisfies the condition. The lattice parameters of  an {\it free-standing} BiTe monolayer  is 4.356 $\AA$ according  to our calculation. In comparison, the lattice constant of the  Bi$_2$Te$_3$(111) surface is 4.335 $\AA$ and 3.755 $\AA$   along $[0\bar 1 1]$  and $[\bar 2 11]$, respectively. Obviously it  is well commensurate along the  $[0\bar 1 1]$ direction of  Bi$_2$Te$_3$ with the BiTe monolayer.
We thus construct the BiTe/Bi$_2$Te$_3$(111) interface  using a slab with a $ (6\times1)$ supercell of Bi$_2$Te$_3$(111) surface which  accommodates atop $(5\times1)$ BiTe supercell, as shown in Fig.\ \ref{fig5}.

The BiTe monolayer after relaxation becomes slightly undulated, with the average distance of 3.71 $\AA$ and 2.79 $\AA$ away from the surface for Te and Bi atoms, respectively.  However, it is worthy to note that the monolayer still keeps the original squared periodicity, consistent with the experimental observation. The formation energy of a BiTe monolayer  on the Bi$_2$Te$_3$(111) surface is calculated to be 0.44 eV per atom under the Te-rich  condition, which is quite smaller than that of an isolated Bi or Te defect atom either on the  surface or in the gap.
It follows that once the diffusion  of the defects is thermally activated, Bi and Te defect atoms tend to aggregate and form a BiTe monolayer.
Note that the BiTe supercell may have three equivalent orientation 120$^\circ$ with each other due to the hexagonal symmetry of the substrate, which is consistent with the experimental observations.
We also construct a slab with a $(4\times1)$ BiTe  supercell on the $ (12\times1)$ supercell of Bi$_2$Te$_3$(111) surface to model the step height of the BiTe monolayer. As shown in Fig.\ \ref{fig2}(b), the obtained 2.3 $\AA$  is well consistent with the experimental result.

It is well-known that a monolayer is  unstable against the formation of domains via elastic relaxation if there is a mismatch between the surface stress of the monolayer and that of the substrate\cite{rmp1125}. The instability wavelength in a specific direction can be roughly estimated as follows \cite{SROLOVITZ1989621,PhysRevLett.61.1973},
\begin{equation}
l=a\exp(\frac{\eta E}{\Delta\sigma^{2}}+1).
\label{eqn:eq5}
\end{equation}
For present system,  $a$ is the lattice constant of the BiTe monolayer,  $\eta$  the edge energy, $E$  the Young modulus, and $\Delta \sigma$ is the stress tensor difference between the Bi$_2$Te$_3$(111) surface with and without the BiTe monolayer. We can assume $\eta \simeq E_{surf}\cdot h $, where $E_{surf}$ and $h$ are the surface energy of Bi$_2$Te$_3$(111) surface and the thickness of the BiTe monolayer, respectively.  According to our calculation, $E = 0.013$ eV/$\AA^{3}$,  $\Delta \sigma_{[\bar 2 11]} = 8.3$ GPa and  $\Delta \sigma_{[0\bar11]} = 5.3$ GPa along the $ [\bar211]$ and $[0\bar11]$ directions, respectively. Thus, we obtain the instability wavelength as 18 and 32 $\AA$ in $ [\bar211]$ and $[0\bar11]$ directions, respectively, which are consistent with the typical length scale of the Mosaic pattern observed experimentally.

In summary, we demonstrate for the first time a post-growth aging effect on Bi$_2$Te$_3$ (111) surface in UHV, i.e. the aggregation of BiTe monolayer on Bi$_2$Te$_3$ (111) surface, which is facilitated by the low formation energy and the thermally activated diffusion of intercalated Bi and Te atoms in the vdW gap. 
In contrast to the previous studies which can be well explained by surface decomposition or reaction, this study uncovers an intrinsic process originated from a novel diffusion process of the defects within the vdW gaps.
 Considering the common nature of weakly bonding between vdW layers, we expect such unusual formation, diffusion and aggregation of defects may be of general importance for the stability and performance of devices based on  most kinds of vdW layered materials.




This work was supported by the Ministry of Science and Technology of China (Grants No.2014CB921103 and No. 2013CB922103),the National Natural Science Foundation of China (Grants No. 11374140, No. 11174123,No. 11474157, No. 11674155 and No. 11634005), the Basic Research Project of Jiangsu Province(Grant No. BK20161390) and the Open Research Fund Program of the State Key Laboratory of Low-Dimensional Quantum Physics. The numerical calculations were carried out at the National Supercomputing Center in Tianjin.

Z.-W. W. and W.-K. H. contributed equally to this work.


\end{document}